\documentclass[a4paper]{jpconf}

\usepackage{graphicx}
\usepackage{amssymb}

\begin{document}
\title{Parametrization for chemical freeze-out conditions from net-charge fluctuations measured at RHIC}

\author{M~Bluhm$^{1}$, P~Alba$^2$, W~Alberico$^2$, R~Bellwied$^3$, V~Mantovani~Sarti$^2$, M~Nahrgang$^4$, C~Ratti$^{2,3}$}

\address{$^1$ Department of Physics, North Carolina State University, Raleigh, NC 27695, USA\\
$^2$ Department of Physics, Torino University and INFN, Sezione di Torino, via P. Giuria 1, 10125 Torino, Italy\\
$^3$ Department of Physics, University of Houston, Houston, TX 77204, USA\\
$^4$ Department of Physics, Duke University, Durham, NC 27708, USA}

\ead{mbluhm@ncsu.edu}

\begin{abstract}
We discuss details of our thermal model applied to extract chemical freeze-out conditions from 
fluctuations in the net-electric charge and net-proton number measured at RHIC. A parametrization for these 
conditions as a function of the beam energy is given. 
\end{abstract}

\section{Introduction}

Over the past decade, statistical hadronization model (SHM) approaches, cf.~e.g.~\cite{Cleymans:2005xv}, 
were quite successful in describing the particle multiplicities measured in heavy-ion collisions at 
various beam energies $\sqrt{s}$, allowing to deduce the chemical freeze-out parameters as unique functions 
of the beam energy. As a complementary method for determining the conditions at freeze-out, fluctuations in the 
conserved charges of QCD, i.e.~baryon number $B$, electric charge $Q$ and strangeness $S$, were 
proposed~\cite{Karsch:2012wm}. The latter can be calculated from first-principles in lattice 
QCD~\cite{Bazavov:2012vg,Borsanyi:2013hza} and compared with corresponding measurements of the moments of 
multiplicity distributions. Experimental data on fluctuations in the net-electric charge and the 
net-proton number (not a conserved charge but often assumed to be a good proxy for net-baryon number) 
became recently available from the beam energy scan program at RHIC~\cite{Adamczyk:2013dal,Adamczyk:2014fia}. 
Thermal models provide another possibility to extract the freeze-out parameters from fluctuation observables. 
In the approach discussed here, we consider a grandcanonical ensemble of hadrons and resonances as 
listed in~\cite{PDG05}. This implies that charge conservation is respected in the means but not in the 
higher-order moments. Given that this as well as other non-included fluctuation sources, e.g.~critical 
fluctuations, have a negligible influence on the observables we study, we show that an extraction of the chemical 
freeze-out parameters is feasible within our approach. 

\section{Freeze-out conditions from moments of net-charge distributions}

In this work, we concentrate on the lowest moments of the measured multiplicity distributions and determine 
the conditions at chemical freeze-out such that the experimental data~\cite{Adamczyk:2013dal,Adamczyk:2014fia} 
for mean $M$ and variance $\sigma^2$ are 
reproduced within error bars. For the net-electric charge one has $M_{{\rm net}-Q}=\sum_i Q_i \langle N_i\rangle$, 
where $\langle N_i\rangle$ is the average of the number $N_i$ of particles of type $i$ with electric charge 
$Q_i$, and $\sigma^2_{{\rm net}-Q}=\sum_i\sum_j Q_iQ_j\langle \Delta N_i\Delta N_j\rangle$ with 
$\Delta N=N-\langle N\rangle$. The sums run over all charged hadrons contained in the measurement. As electric 
charge is dominated by $\pi^+$, $\pi^-$, $K^+$, $K^-$, $p$ and $\bar{p}$, we restrict our model to this ensemble. 
For the net-proton number one has $M_{p-\bar{p}}=\langle N_p\rangle-\langle N_{\bar{p}}\rangle$ and 
$\sigma^2_{p-\bar{p}}=\langle (\Delta N_p)^2\rangle+\langle (\Delta N_{\bar{p}})^2\rangle-
2\langle \Delta N_p\Delta N_{\bar{p}}\rangle$, where the covariance between protons and anti-protons 
$\langle \Delta N_p\Delta N_{\bar{p}}\rangle$ vanishes 
in our approach. To take the actual experimental situation more accurately into account, several physically 
relevant aspects are included in our study as explained in the following. 

\subsection{Influence of kinematic cuts and resonance decays}

The experimental coverage in rapidity (or pseudo-rapidity) and transverse momentum is limited by the detector 
design and the demands from particle reconstruction and identification. To account for this acceptance limitation, 
it was proposed in~\cite{Garg:2013ata} to restrict the phase-space integrals in a thermal model correspondingly. 
In~\cite{Garg:2013ata}, the sensitivity of moment ratios to kinematic cuts was found to be small for net-protons 
and more pronounced for the net-electric charge. We follow this proposal and apply kinematic cuts in line with 
the experimental analysis~\cite{Adamczyk:2013dal,Adamczyk:2014fia}. 

Resonance decays can, event by event, significantly influence the final multiplicities of stable (with respect to 
strong and electromagnetic decays) particles. A contamination from secondaries, e.g.~weak decay contributions 
from $\Lambda$ to $p$ or spallation protons from the beam pipe, was reported to be suppressed through the 
experimental analysis~\cite{Adamczyk:2013dal}. In our model, contributions from resonance decays to the averages 
$\langle N_i\rangle$ and $\langle\Delta N_i\Delta N_j\rangle$ in $M$ and $\sigma^2$ are included via 
\begin{eqnarray}
\label{equ:1}
 \langle N_i\rangle & = & \langle N_i^*\rangle_T+\sum_R \langle N_R^*\rangle_T \langle n_i\rangle_R \,, \\
\label{equ:2}
 \langle\Delta N_i\Delta N_j\rangle & = & \langle\Delta N_i^*\Delta N_j^*\rangle_T + 
 \sum_R \langle(\Delta N_R^*)^2\rangle_T \langle n_i\rangle_R \langle n_j\rangle_R + 
 \sum_R \langle N_R^*\rangle_T \langle \Delta n_i\Delta n_j\rangle_R \,.
\end{eqnarray}
In Eqs.~(\ref{equ:1}) and~(\ref{equ:2}), $N_i^*$ and $N_R^*$ denote the primordial (before resonance decay) 
numbers of stable particles $i$ and resonances $R$, while $\langle n_i\rangle_R=\sum_r b_r^R n_{i,r}^R$ and 
$\langle \Delta n_i\Delta n_j\rangle_R=\sum_r b_r^R n_{i,r}^R n_{j,r}^R-\sum_r b_r^R n_{i,r}^R\cdot
\sum_{r'} b_{r'}^R n_{j,r'}^R$ are decay-channel averages, with $r$-th channel branching ratio $b_r^R$ and 
$n_{i,r}^R$ the number of $i$ produced in that decay-channel of $R$. As the decay of a resonance is a 
probabilistic process, causing itself event-by-event fluctuations in the final hadron multiplicities, the 
averages in $M$ and $\sigma^2$ involve an averaging over the thermal ensemble and over resonance decays. 
The thermal, primordial means and (co)variances are obtained in our model by derivatives of the 
pressure $P$ with respect to the particle chemical potentials $\mu_l$ as (indices $i$ and $j$ stand for 
hadrons or resonances here) 
\begin{equation}
\label{equ:3}
 \langle N_i^*\rangle_T = VT^3\, \frac{\partial(P(T,\{\mu_l\})/T^4)}{\partial(\mu_i/T)} \,, \quad
 \langle \Delta N_i^*\Delta N_j^*\rangle_T = VT^3\, \frac{\partial^2(P(T,\{\mu_l\})/T^4)}{\partial(\mu_i/T)\partial(\mu_j/T)} \,.
\end{equation}

Although in the grandcanonical ensemble the primordial covariances $\langle\Delta N_i^*\Delta N_j^*\rangle_T$ 
between different particle species vanish, resonance decays can induce correlations such that the covariances between 
different hadrons affect the moments of the net-electric charge distribution. Due to baryon-number 
conservation this is, however, not the case for net protons. Interestingly, the form of the factors 
$\langle \Delta n_i\Delta n_j\rangle_R$ implies that the fluctuation contributions associated with the probabilistic 
nature of resonance decays can be small or zero depending on the charge under consideration: if the 
number of produced charges is the same in each decay channel of $R$, they will exactly vanish. In the case of the 
net-electric charge this means that although individual factors such as $\langle \Delta n_{\pi^+}\Delta n_{K^-}\rangle_R$ 
can be non-zero, the probabilistic contributions cancel each other exactly in the double-sum over all charged hadrons 
in $\sigma^2_{{\rm net}-Q}$ because electric charge is conserved in each decay. Since we restrict the ensemble of 
considered hadrons, however, we expect some but sub-dominant contributions from resonances decaying, in addition, into 
charged hyperons. In the case of the net-proton number, probabilistic fluctuation contributions are significant and 
vanish only for the $\Delta^{++}$-resonances. A systematic study for net protons was presented in~\cite{Nahrgang:2014fza}. 

\subsection{Influence of the isospin randomization of nucleons}

Final state effects may significantly influence fluctuations, in particular, in the net-proton number. The 
regeneration and subsequent decay of $\Delta$-resonances via 
$p(n)+\pi\to\Delta\to n(p)+\pi$ can change the isospin identity of the nucleons $p$ 
and $n$ (similar for anti-nucleons) and, thus, influence their multiplicity distributions. As the electric charge is 
conserved in these reactions, they do not affect the fluctuations in the net-electric charge. Also, the mean 
$M_{p-\bar{p}}$ is not altered, but the higher-order moments of the net-proton distribution are influenced. 
For high enough pion densities and a long enough hadronic phase (before free streaming), such 
that a full isospin randomization can be assumed, correction expressions for the moments of the net-proton distribution 
based on the moments of baryon and anti-baryon distributions can be derived, cf.~\cite{Kitazawa:2011wh,Kitazawa:2012at}. 
It was argued in~\cite{Kitazawa:2011wh,Kitazawa:2012at} that the necessary conditions are satisfied for 
$\sqrt{s}\gtrsim 10$~GeV. 

Excluding explicitly weak decay contributions from hyperons, these correction expressions may be formulated in 
terms of the moments of the nucleon and anti-nucleon distributions. Accordingly, the effect of isospin randomization 
is included in our model as 
\begin{equation}
\label{equ:4}
 \sigma^2_{p-\bar{p}} = \langle (r\cdot\Delta N_N-\bar{r}\cdot\Delta N_{\bar{N}})^2\rangle + 
 \langle r(1-r)N_N+\bar{r}(1-\bar{r})N_{\bar{N}})\rangle
\end{equation}
with $r=\langle N_p\rangle/\langle N_N\rangle$, $\bar{r}=\langle N_{\bar{p}}\rangle/\langle N_{\bar{N}}\rangle$ 
and $N_N=N_p+N_n$, $N_{\bar{N}}=N_{\bar{p}}+N_{\bar{n}}$. The entering quantities are calculated via 
Eqs.~(\ref{equ:1}) -~(\ref{equ:3}). As shown in~\cite{Nahrgang:2014fza}, the main effect of these corrections is to 
make the net-proton distribution look Poissonian independent of whether the probabilistic fluctuation contributions 
from resonance decays are taken into account or not. 

\section{Parametrization for the chemical freeze-out conditions}

With the abovementioned refinements in our thermal model, one can determine the values for temperature $T$ and 
baryo-chemical potential $\mu_B$ at chemical freeze-out by comparing the measured moment ratios 
$\sigma^2_{{\rm net}-Q}/M_{{\rm net}-Q}$ and $\sigma^2_{p-\bar{p}}/M_{p-\bar{p}}$ with the model results, 
fulfilling additionally physical constraints on $M_{{\rm net}-Q}/M_{{\rm net}-B}$ and 
$M_{{\rm net}-S}/M_{{\rm net}-B}$. In the moment ratios, the volume $V$ appearing in Eq.~(\ref{equ:3}) cancels 
when ignoring fluctuations in $V$ itself. The corresponding parameter values for $T$ and $\mu_B$ were reported 
in~\cite{Alba:2014eba} and contrasted with different results from lattice QCD and with SHM results 
in~\cite{Bluhm:2014wha}. 

\begin{figure}[t]
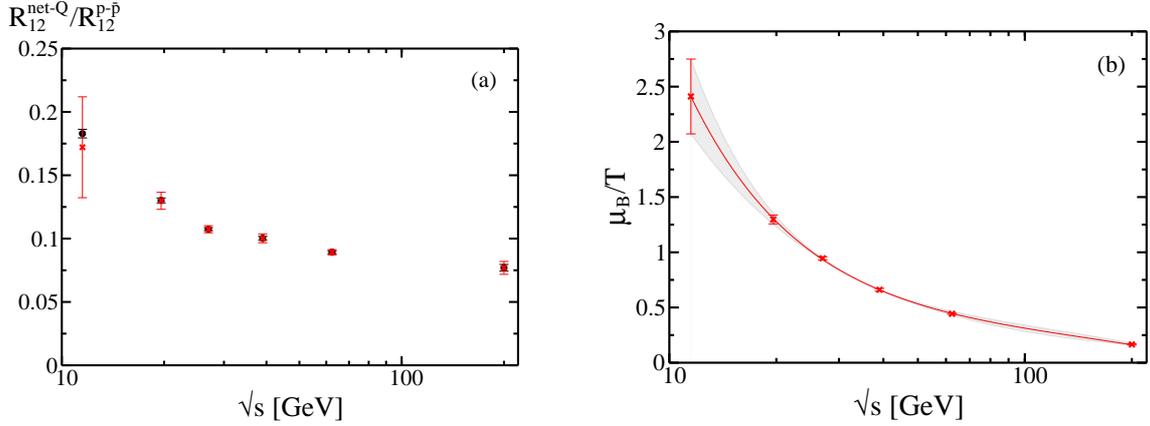

\begin{center}
\includegraphics[width=16pc]{R12Comparison.eps}
\hspace{2pc}
\includegraphics[width=17pc]{muBTRatio.eps}
\end{center}
\caption{\label{Fig:1}(Color online) (a): double-ratio of measured $\sigma^2/M$-ratios for net protons and 
net-electric charge~\cite{Adamczyk:2013dal,Adamczyk:2014fia} (circles) as a function of $\sqrt{s}$, compared 
with our thermal model results (crosses) using the chemical freeze-out parameters from~\cite{Alba:2014eba}. (b): ratio 
$\mu_B/T$ of these freeze-out parameters (crosses) as a function of $\sqrt{s}$ together with a suitable 
parametrization thereof (solid curve) of the form Eq.~(\ref{equ:5}) with parameter values 
$a_0=57.24\pm^{73.78}_{35.91}$, $a_1=1.345\pm^{0.298}_{0.386}$, $a_2=0.276\pm^{0.116}_{0.251}$ and 
$a_3=-0.00080\mp^{0.00040}_{0.00082}$ (errors define the shaded band).}
\end{figure}

In the left panel of Fig.~\ref{Fig:1} we show the quality of the description of the double-ratio 
$R_{12}^{{\rm net}-Q}/R_{12}^{p-\bar{p}}=(\sigma^2_{p-\bar{p}}/M_{p-\bar{p}})/(\sigma^2_{{\rm net}-Q}/M_{{\rm net}-Q})$ 
of the measured moment ratios with our model obtained for the freeze-out parameters~\cite{Alba:2014eba}. Since we assume 
full isospin randomization independent of the beam energy, we restrict our study to $\sqrt{s}\geq 11.5$~GeV. 
The ratio $\mu_B/T$ of our freeze-out parameters (with error bars based on the reported errors in the experimental 
data~\cite{Adamczyk:2013dal,Adamczyk:2014fia}) is shown in the right panel of Fig.~\ref{Fig:1} (symbols). A 
suitable parametrization of these results as a function of $\sqrt{s}$ 
within $11.5\leq\sqrt{s}/$GeV~$\leq 200$ can be given in the form 
\begin{equation}
\label{equ:5}
 \frac{\mu_B}{T} = \frac{a_0}{(\sqrt{s}/{\rm GeV})^{a_1}}+a_2+a_3(\sqrt{s}/{\rm GeV}) 
\end{equation}
with $a_0=57.24$, $a_1=1.345$, $a_2=0.276$ and $a_3=-0.00080$, which describes accurately our central values 
for $\mu_B/T$ (cf.~solid curve in the right panel of Fig.~\ref{Fig:1}). 

\section{Conclusion}

We described in detail the physically relevant aspects taken into account in our thermal model in order to 
successfully deduce chemical freeze-out parameters from comparing model results with the moment ratios 
$\sigma^2/M$ of the measured net-electric charge and net-proton distributions. We provided a suitable 
parametrization for the ratio $\mu_B/T$ of extracted freeze-out conditions as a function of $\sqrt{s}$. As 
discussed in~\cite{Bluhm:2014wha}, other sources of fluctuations that are not included in our approach, 
such as critical fluctuations, volume fluctuations or the influence of exact charge conservation, have a 
negligible impact on these results for the highest beam energies considered in our analysis. 

\ack
This work is supported by the Italian Ministry of Education, Universities and Research under the FIRB 
Research Grant RBFR0814TT, by the DAAD and by the U.S. Department of Energy grants DE-FG02-03ER41260, 
DE-FG02-05ER41367 and DE-FG02-07ER41521.

\end{document}